\documentclass[showpacs,amsmath,amssymb,nofootinbib,prd]{revtex4}
\newcommand{\Tr}{\operatorname{Tr}}
\begin{document}

\title{Renormalization Group Equations for the CKM matrix}

\author{P.~Kielanowski} \email{kiel@physics.utexas.edu}
\affiliation{Departamento de F\'{\i}sica, Centro de
Investigaci\'{o}n
  y Estudios Avanzados del IPN, Mexico}

\author{S.R.~Ju\'{a}rez~W.} \email{rebeca@esfm.ipn.mx}
\affiliation{Departamento de F\'{\i}sica, Escuela Superior de
  F\'{\i}sica y Matem\'{a}ticas, IPN, Mexico}

\author{J.H. Montes de Oca Y.} \email{josehalim@hotmail.com}
\affiliation{Departamento de F\'{\i}sica, Escuela Superior de
  F\'{\i}sica y Matem\'{a}ticas, IPN, Mexico}

\begin{abstract}
  We derive the one loop renormalization group equations for the
  Cabibbo-Kobayashi-Maskawa matrix for the Standard Model, its two
  Higgs extension and the minimal supersymmetric extension in a novel
  way. The derived equations depend only on a subset of the model
  parameters of the renormalization group equations for the quark
  Yukawa couplings so the CKM matrix evolution cannot fully test the
  renormalization group evolution of the quark Yukawa couplings. From
  the derived equations we obtain the invariant of the renormalization
  group evolution for three models which is the angle $\alpha$ of the
  unitarity triangle. For the special case of the Standard Model and
  its extensions with $v_{1}\approx v_{2}$ we demonstrate that also
  the shape of the unitarity triangle and the Buras-Wolfenstein
  parameters $\bar{\rho}=\left(1-\frac{1}{2}\lambda^{2}\right)\rho$
  and $\bar{\eta}=\left(1-\frac{1}{2}\lambda^{2}\right)\eta$ are
  conserved. The invariance of the angles of the unitarity triangle
  means that it is not possible to find a model in which the CKM
  matrix might have a simple, special form at asymptotic energies.
\end{abstract}

\maketitle

\section{Introduction}
To obtain finite results in quantum field theory, in a higher order
than the tree level, one has to perform the renormalization program.
The independence of the renormalization procedure of the
renormalization point leads to the dependence of the Lagrangian
parameters on the point of renormalization. This dependence is
governed by the renormalization group equations for the coupling
constants and other parameters of the
Lagrangian~\cite{HelvPhysActa.26.499,PhysRev.95.1300}. The most
important application of the renormalization group equations is a
possibility to study the asymptotic properties of the Lagrangian
parameters like the running masses and the coupling constants (for a
very early example see \textit{e.g.},~\cite{PhysRevLett.33.451}). The
theories at the asymptotic energies may reveal some new symmetries or
other interesting properties that give a deeper insight into the
physical content.  Also from the physical requirement of the stability
of the theory one can \textit{e.g.}, estimate the range of the
physical parameters like the Higgs mass in the Standard Model or its
extensions~\cite{kielanowski:096003} and the references therein.

The observable parameters of the Standard Model are the following: 6
quark masses, 3 lepton masses, 4 parameters of the
Cabibbo--Kobayashi--Maskawa (CKM) matrix~\cite{Kobayashi,Cabibbo} and
3 gauge coupling constants. The CKM matrix is obtained from the
diagonalizing matrices of the Yukawa couplings.

The renormalization group (RG) equations for the CKM matrix is
obtained from the renormalization group equations for the Yukawa
couplings. In this paper we derive the RG equations for the CKM matrix
for the Standard Model, two Higgs doublet extension of the Standard
Model and the Minimal Supersymmetric Standard Model. We first get a
complete one loop equations without any approximation and next discuss
these equations taking into account the hierarchy contained in the
Yukawa couplings and the CKM matrix elements.

The interesting fact that can be observed is that the equations for
the CKM matrix do not depend on all the parameters of the original
equations for the Yukawa couplings. Finally we discuss the physical
implications of the derived equations.

\section{RG equations for the CKM matrix}\label{RGeq}
\subsection{Preliminary considerations}
The part of the Lagrangian of the Standard Model and its extensions
that contains the Yukawa interactions of the quarks has the following
structure
\begin{equation}
  \mathcal{L}_{Y}=y_{u}\overline{q}_{L}\Phi
  _{1}(q_{u})_{R}+y_{d}\overline{q}_{L}{\Phi
  }_{2}(q_{d})_{R}+\text{h.c.}  \label{1}
\end{equation}
Here the $y_{u}$ and the $y_{d}$ are 3$\times $3 matrices of the
Yukawa coupling of the up and down quarks, $\Phi_{1}$ and $\Phi_{2}$
are the two Higgs doublets (in case of the Standard Model
$\Phi_{2}=\Phi$, $\Phi_{1}=\epsilon\Phi^{*}$, $\epsilon$ is the
$2\times 2$ antisymmetric tensor). The $q_{L}$ and ($q_{u,d}$)$_{R}$
are the left-handed quark doublet and right quark singlets,
respectively.  Eq.~\eqref{1} implicitly contains the summation of the
quark generations. The Yukawa couplings can be diagonalized with the
help of the biunitary transformation:
\begin{equation}
  y_{u}=U_{L}^{\dagger
  }Y_{u}U_{R},\quad\quad y_{d}=D_{L}^{\dagger }Y_{d}D_{R}\;,  \label{2} 
\end{equation}
where $U_{L}$, $U_{R}$, $D_{L}$, $D_{R}$ are unitary, diagonalizing
matrices and $Y_{u}$ and $Y_{d}$ are the diagonal matrices with
positive diagonal elements. The quark masses (running) are equal to
\begin{equation}
  \begin{array}{lll}
  \displaystyle m_{u} =\frac{1}{\sqrt{2}}v_{1}\left( Y_{u}\right) _{11},\quad
  &\displaystyle m_{c}=\frac{1}{\sqrt{2}}v_{1}\left( Y_{u}\right) _{22},\quad
  &\displaystyle m_{t}=\frac{1}{\sqrt{2}} v_{1}\left( Y_{u}\right)
  _{33},\\ 
  \displaystyle m_{d} =\frac{1}{\sqrt{2}}v_{2}\left( Y_{d}\right) _{11},\quad 
  &\displaystyle m_{s}=\frac{1}{\sqrt{2}}v_{2}\left( Y_{d}\right)_{22},\quad
  &\displaystyle m_{b}=\frac{1}{\sqrt{2}}v_{2}\left( Y_{d}\right)_{33}.
\end{array} \label{3} 
\end{equation}
Here $v_{1}$ and $v_{2}$ are the vacuum expectation values of the
Higgs fields and in the Standard Model $v_{1}=v_{2}=v$.

From Eqs.~(\ref{3}) one can see that there is a hierarchy between the
elements of $(Y_{u})_{ii}$ and bewteen the elements of $(Y_{d})_{ii}$
that follows from the hierarchy of the quark masses.

The CKM matrix of the charged weak current is obtained from the left
diagonalizing matrices $U_{L}$ and $D_{L}$~\footnote{ We use the
  following notation
\begin{equation*}
\left(
\begin{array}{ccc}
    V_{11},&V_{12},&V_{13}\\
    V_{21},&V_{22},&V_{23}\\
    V_{31},&V_{32},&V_{33}
  \end{array}\right)
  =\left(\begin{array}{ccc}
    V_{ud},&V_{us},&V_{ub}\\
    V_{cd},&V_{cs},&V_{cb}\\
    V_{td},&V_{ts},&V_{tb}
\end{array}\right).
\end{equation*}
} 
\begin{equation}
V=U_{L}D_{L}^{\dagger }.  \label{4}
\end{equation}
The matrices $U_{R}$ and $D_{R}$ are not related to any observables in
absence of the right handed currents.

The general structure of the one loop renormalization group equations
for the Standard Model and its extensions for the Yukawa couplings
$y_{u}$ and $y_{d}$ is the
following~\cite{PhysRevD.46.3945,PhysRevD.47.232}
\begin{subequations}
\label{5}
\begin{eqnarray}
\frac{dy_{u}}{dt} &=&\frac{1}{(4\pi )^{2}}[\alpha _{1}^{u}(t)+\alpha
_{2}^{u}y_{u}^{\phantom {\dagger }}y_{u}^{\dagger }+\alpha _{3}^{u}{\Tr}
(y_{u}^{\phantom {\dagger }}y_{u}^{\dagger })+\alpha _{4}^{u}(y_{d}^{
\vphantom{\dagger}}y_{d}^{\dagger })+\alpha _{5}^{u}{\Tr}(y_{d}^{
\vphantom{\dagger}}y_{d}^{\dagger })]y_{u},  \label{5a} \\
\frac{dy_{d}}{dt} &=&\frac{1}{(4\pi )^{2}}[\alpha _{1}^{d}(t)+\alpha
_{2}^{d}y_{u}^{\phantom {\dagger }}y_{u}^{\dagger }+\alpha _{3}^{d}{\Tr}
(y_{u}^{\phantom {\dagger }}y_{u}^{\dagger })+\alpha _{4}^{d}(y_{d}^{
\vphantom{\dagger}}y_{d}^{\dagger })+\alpha _{5}^{d}{\Tr}(y_{d}^{
\vphantom{\dagger}}y_{d}^{\dagger })]y_{d}.  \label{5b}
\end{eqnarray}
\end{subequations}
Here $t=\ln(\mu/m_{t})$, $\mu$ is the energy of the
renormalization point and $m_{t}$ is the top quark mass.
Eqs.~(\ref{5}) form a system of coupled non-linear equations. The
parameters $\alpha _{1}^{u}(t),\,\alpha _{1}^{d}(t)$ are quadratic
functions of the gauge coupling constants and $\alpha_{i}^{u}$,
$\alpha _{i}^{d}$, $i=2,\ldots,5$ are constants.  The values of the
$\alpha _{i}^{u}(t)$ and $\alpha _{i}^{d}(t)$ do depend on the model
and they are given in Appendix.

\subsection{Derivation of the RG equations for the CKM matrix}

We will now obtain from Eqs.~\eqref{5} the renormalization group
equations for the elements of the CKM matrix (\ref{4}). The derivation
is done in the following steps:
\begin{enumerate}
\item Insert representation (\ref{2}) in Eqs.~(\ref{5}).
\item Eliminate the matrices $U_{R}$ and $D_{R}$ from the equations.
\item Derive the final form of the equations for the CKM matrix.
\end{enumerate}
Before the derivation, let us observe that after the differentiation
of the unitarity relations $U^{\dagger }U=1$ and $UU^{\dagger }=1$ we
obtain
\begin{subequations}
\label{6}
\begin{eqnarray}
&&\frac{dU^{\dagger }}{dt}U+U^{\dagger }\frac{dU}{dt} =\left( \frac{
dU^{\dagger }}{dt}U\right) +\left( \frac{dU^{\dagger }}{dt}U\right)
^{\dagger }=\left(U^{\dagger }\frac{dU}{dt}\right) +\left(U^{\dagger}
\frac{dU}{dt}\right)^{\dagger}=0  \label{6a} \\
&&\frac{dU}{dt}U^{\dagger }+U\frac{dU^{\dagger }}{dt} =\left( \frac{dU}{dt}
U^{\dagger }\right) +\left( \frac{dU}{dt}U^{\dagger }\right) ^{\dagger
} =\left(U\frac{dU^{\dagger }}{dt} \right) +\left(U\frac{dU^{\dagger
    }}{dt} \right)^{\dagger} =0 
\label{6b}
\end{eqnarray}  
\end{subequations}
\textit{i.e.}, the matrices $ \frac{dU^{\dagger }}{dt}U$,
$U^{\dagger }\frac{dU}{dt}$, $\frac{dU}{ dt}U^{\dagger }$
and $U\frac{dU^{\dagger }}{dt}$ are antihermitian.

In the derivation of the RG equations for the CKM matrix we will
explicitly show all the algebraic operations that are necessary for
Eq.~(\ref{5a}), the analogous steps for Eq.~(\ref{5b}) are not shown
and only the final results are quoted.

\noindent
\textbf{Step 1}\\
From Eq.~\eqref{2} and~\eqref{5a} we obtain
\begin{equation}
  \frac{d\left( U_{L}^{\dagger }Y_{u}U_{R}\right) }{dt}
  =\frac{1}{(4\pi )^{2}}\left(\alpha _{1}^{u}(t)+\alpha _{2}^{u}U_{L}^{\dagger
  }Y_{u}^{\,\,\,2}U_{L}+\alpha _{3}^{u}{\Tr}(Y_{u}^{\,\,\,2})+\alpha
  _{4}^{u}(\,\,D_{L}^{\dagger }Y_{d}^{\,\,\,2}D_{L})+\alpha _{5}^{u}{\Tr}
  (Y_{d}^{\,\,\,2})\right)U_{L}^{\dagger }Y_{u}U_{R}.  \label{7}
\end{equation}
After differentiation, the left hand side of Eq.~(\ref{7}) becomes
\begin{equation}
\frac{dU_{L}^{\dagger }}{dt}Y_{u}U_{R}+U_{L}^{\dagger }\frac{dY_{u}}{dt}
U_{R}+U_{L}^{\dagger }Y_{u}\frac{dU_{R}}{dt}.  \label{8}
\end{equation}

\noindent
\textbf{Step 2}\\
Using relation~\eqref{7} we multiply Eq.~\eqref{8} from the left by
$U_{L}$ and from the right by $U_{R}^{\dagger }Y_{u}$
\begin{equation}
  U_{L}\frac{dU_{L}^{\dagger }}{dt}Y_{u}^{\,\,\,2}+\frac{dY_{u}}{dt}
  Y_{u}+Y_{u}\frac{dU_{R}}{dt}U_{R}^{\dagger }Y_{u}
  =\frac{1}{(4\pi )^{2}}\left(\alpha _{1}^{u}(t)+\alpha
  _{2}^{u}Y_{u}^{\,\,\,2}+\alpha _{3}^{u}{\Tr}(Y_{u}^{\,\,\,2})+\alpha
  _{4}^{u}\,VY_{d}^{\,\,\,2}V^{\dagger }+\alpha _{5}^{u}{\Tr}
  (Y_{d}^{\,\,\,2})\right)Y_{u}^{\,\,\,2}  \label{9}
\end{equation}
and next we add to Eq.~\eqref{9} its hermitian conjugate. Then from
the property~(\ref{6b}) the term with the $U_{R}$ matrix cancels out
and we obtain
\begin{subequations}
\label{10}
\begin{equation}
\frac{dY_{u}^{\,\,\,2}}{dt}+\left[ U_{L}\frac{dU_{L}^{\dagger }}{dt}
,Y_{u}^{\,\,\,2}\right]
=\frac{1}{(4\pi )^{2}}\left( 2\left( \alpha _{1}^{u}(t)+\alpha
_{2}^{u}Y_{u}^{\,\,\,2}+\alpha _{3}^{u}{\Tr}(Y_{u}^{\,\,\,2})+\alpha
_{5}^{u}{\Tr}(Y_{d}^{\,\,\,2})\right) Y_{u}^{\,\,\,2}+\alpha
_{4}^{u}\left\{ VY_{d}^{\,\,\,2}V^{\dagger },Y_{u}^{\,\,\,2}\right\} \right).
\label{10a}
\end{equation}
Here $[\;\cdot\; ,\;\cdot\;] $ and $\{\; \cdot\; ,\;\cdot\;\} $ are
the commutator and anticommutator of two matrices.  From
Eq.~(\ref{5b}) we get
\begin{equation}
  \frac{dY_{d}^{\,\,\,2}}{dt}+\left[ D_{L}\frac{dD_{L}^{\dagger }}{dt}
    ,Y_{d}^{\,\,\,2}\right]
  =\frac{1}{(4\pi )^{2}}\left( 2\left( \alpha _{1}^{d}(t)+\alpha _{3}^{d}
      {\Tr}(Y_{u}^{\,\,\,2})+\alpha _{4}^{d}Y_{d}^{\,\,\,2}+\alpha _{5}^{d}
      {\Tr}(Y_{d}^{\,\,\,2})\right) Y_{d}^{\,\,\,2}+\alpha _{2}^{d}\left\{
      V^{\dagger }Y_{u}^{\,\,\,2}V,Y_{d}^{\,\,\,2}\right\} \right).  \label{10b}
\end{equation}
\end{subequations}
Eqs.~(\ref{10}) are matrix type differential equations of dimension
$3\times 3$. The equations for the diagonal elements give the
renormalization group equations for the diagonal elements $\left(
  Y_{u}\right) _{ii}$ and $\left( Y_{d}\right) _{ii}.$ We will not
consider these equations here. The non-diagonal Eqs.~\eqref{10} can be
written in the following form
\begin{subequations}
\label{11}  
\begin{eqnarray}
&&\left( U_{L}\frac{dU_{L}^{\dagger }}{dt}\right) _{ij}=\frac{1}{(4\pi )^{2}}
\alpha _{4}^{u}H_{ij}^{u}\left( VY_{d}^{\,\,\,2}V^{\dagger }\right)
_{ij}\equiv R_{ij}^{u}\,,\quad i\neq j,  \label{11a}\\
&&\left( D_{L}\frac{dD_{L}^{\dagger }}{dt}\right) _{ij}=\frac{1}{(4\pi )^{2}}
\alpha _{2}^{d}H_{ij}^{d}\left( V^{\dagger }Y_{u}^{\,\,\,2}V\right)
_{ij}\equiv R_{ij}^{d}\,,\quad i\neq j,  \label{11b}
\end{eqnarray}
\end{subequations}
with no summation over $i,j$. The matrices $R_{ij}^{u,d}$ are the
right hand sides of Eqs.~(\ref{11}) and $H_{ij}^{u,d}$ are equal to
\begin{equation}
H_{ij}^{u,d}=\frac{\left( Y_{u,d}\right) _{jj}+\left( Y_{u,d}\right) _{ii}}{
\left( Y_{u,d}\right) _{jj}-\left( Y_{u,d}\right) _{ii}}.  \label{12}
\end{equation}
The right hand side of Eqs.~(\ref{11}) is antihermitian, this
guarantees the unitary evolution of the matrices $U_{L}$ and $D_{L}$.

\noindent
\textbf{Step 3}\\
Now we derive the renormalization group equations for the elements of
the CKM matrix $V$.

With Eqs.~(\ref{11}) we obtain
\begin{equation}
  \frac{dU_{L}}{dt}D_{L}^{\dagger }=-R^{u}\cdot V,\quad 
  U_{L}\frac{dD_{L}^{\dagger }}{dt}=V\cdot R^{d}.  \label{13}
\end{equation}
After taking the sum of two equations in~(\ref{13}) we obtain
\begin{equation}
  \frac{dV}{dt}=\frac{dU_{L}}{dt}D_{L}^{\dagger }+U_{L}\frac{dD_{L}^{\dagger }
  }{dt}=-R^{u}\cdot V+V\cdot R^{d}  \label{14}
\end{equation}
or written explicitly in terms matrix elements with indices
\begin{equation}
  \left( \frac{dV}{dt}\right) _{ik}=-R_{ii}^{u}V_{ik}-\sum_{j\neq
    i}R_{ij}^{u}V_{jk}+V_{ik}R_{kk}^{d}+\sum_{j\neq
    k}V_{ij}R_{jk}^{d}. \label{14a}
\end{equation}
Eqs.~(\ref{14}) and~\eqref{14a} are the RG equations for the CKM
matrix, but they are not yet in the final form, because as seen from
Eq.~\eqref{14a} they contain the diagonal elements of the matrix
$\left( R^{u,d}\right) _{ii}$, which cannot be derived from
Eqs.~(\ref{11}), they are unknown and Eq.~(\ref{14}) cannot be
directly solved. It turns out, however, that the problem of unwanted
terms can be overcome for the absolute values of the matrix elements
of the CKM matrix which we will obtain in the later part of the paper
for the approximate equations. The exact form of these equations is
not very useful.

\section{Approximations}\label{approx}

\subsection{Hierarchy of the Yukawa couplings}

Let us notice now that Eqs.~\eqref{14} and~\eqref{14a} are one loop
equations \textit{i.e.}, the two loop and higher order loop terms are
neglected. The loop order parameter of the RG equations is
$1/(4\pi)^{2}\sim\lambda^{4}$, $\lambda \sim \left| V_{12}\right| $,
so it means that we have to keep only the terms of the relative order
lower than $\lambda^{4}$ in Eqs.~\eqref{14} and~\eqref{14a}.  We have
the following hierarchy of the Yukawa couplings that follows from
Eq.~\eqref{3} and the experimental values of the quark
masses~\cite{PDG2008}
\begin{equation}
  \begin{array}{ll}
    \displaystyle
    \frac{\left( Y_{u}^{\,\,\,2}\right) _{11}}{\left( Y_{u}^{\,\,\,2}\right)
      _{33}} =\left( \frac{m_{u}}{m_{t}}\right) ^{2}\sim
    \lambda^{16},\quad\quad\quad
    &\displaystyle\frac{\left( Y_{u}^{\,\,\,2}\right) _{22}}{\left(
        Y_{u}^{\,\,\,2}\right) 
      _{33}}=\left( \frac{m_{c}}{m_{t}}\right) ^{2}\sim \lambda ^{8}, \\
    \displaystyle\frac{\left( Y_{d}^{\,\,\,2}\right) _{11}}{\left(
        Y_{d}^{\,\,\,2}\right) 
      _{33}} =\left( \frac{m_{d}}{m_{b}}\right) ^{2}\sim \lambda^{8},
    &\displaystyle\frac{\left( Y_{d}^{\,\,\,2}\right) _{22}}{\left(
        Y_{d}^{\,\,\,2}\right) 
      _{33}}=\left( \frac{m_{s}}{m_{b}}\right) ^{2}\sim \lambda ^{4}.
  \end{array} \label{21}
\end{equation}

Using the one loop approximation and Eq.~\eqref{21} we have
\begin{equation}
H_{ij}^{u,d}=\begin{cases}

-1& \text{for }i>j\\

+1& \text{for }i<j\\
\text{not defined}& \text{for }i=j
\end{cases}
\label{17}
\end{equation}
and Eq.~\eqref{14a} can be written in the following form\footnote{We
  explicitly write the RG equations for $V_{11}$, $V_{13}$, $V_{31}$
  and $V_{33}$, because these equations are sufficient for the
  determination of the evolution of all the observables of the CKM
  matrix, which are determined by 4~parameters. The equations for the
  5 remaining elements can also be derived, but they are more
  complicated and we do not consider them here.}
\begin{subequations}
  \label{14c}
\begin{eqnarray}
  \label{14ca}
  &&\frac{1}{ V_{11}}\frac{dV_{11}}{dt} = (R_{11}^{d}-R_{11}^{u})
  -\frac{1}{(4\pi )^{2}}
  \alpha_{4}^{u}\left((Y_{d}^{2})_{11}-(VY_{d}^{2}V^{\dagger})_{11}\right)
 -\frac{1}{(4\pi )^{2}}
  \alpha_{2}^{d}\left((Y_{u}^{2})_{11}-(V^{\dagger}Y_{u}^{2}V)_{11}\right),\\
\label{14cb}
  &&\frac{1}{ V_{33}}\frac{dV_{33}}{dt} = (R_{33}^{d}-R_{33}^{u})
   +\frac{1}{(4\pi )^{2}}
  \alpha_{4}^{u}\left((Y_{d}^{2})_{33}-(VY_{d}^{2}V^{\dagger})_{33}\right)
 +\frac{1}{(4\pi )^{2}}
  \alpha_{2}^{d}\left((Y_{u}^{2})_{33}-(V^{\dagger}Y_{u}^{2}V)_{33}\right),\\
\label{14cc}
  &&\frac{1}{ V_{13}}\frac{dV_{13}}{dt} = (R_{33}^{d}-R_{11}^{u})
  -\frac{1}{(4\pi )^{2}}
  \alpha_{4}^{u}\left((Y_{d}^{2})_{33}-(VY_{d}^{2}V^{\dagger})_{11}\right)
 +\frac{1}{(4\pi )^{2}}
  \alpha_{2}^{d}\left((Y_{u}^{2})_{11}-(V^{\dagger}Y_{u}^{2}V)_{33}\right),\\
\label{14cd}
  &&\frac{1}{ V_{31}}\frac{dV_{31}}{dt} = (R_{11}^{d}-R_{33}^{u})
  +\frac{1}{(4\pi )^{2}}
  \alpha_{4}^{u}\left((Y_{d}^{2})_{11}-(VY_{d}^{2}V^{\dagger})_{33}\right)
 -\frac{1}{(4\pi )^{2}}
  \alpha_{2}^{d}\left((Y_{u}^{2})_{33}-(V^{\dagger}Y_{u}^{2}V)_{11}\right).
\end{eqnarray}
\end{subequations}

\subsection{Constants of the RG evolution}

Let us notice that in Eqs.~\eqref{14c} on the right hand side the
terms $R^{u,d}_{ii}$ are purely imaginary and the rest of the terms
are real. From this we can obtain the quantities that do not evolve
with energy. Observing that the following expressions are purely real
\begin{subequations}
    \label{14d}
  \begin{eqnarray}
    \label{14da}
    &&\frac{1}{ V_{11}}\frac{dV_{11}}{dt} - \left(\frac{1}{
      V_{33}}\frac{dV_{33}}{dt}\right)^{*} 
  +\left( \frac{1}{ V_{13}}\frac{dV_{13}}{dt}\right)^{*}
    -\frac{1}{ V_{31}}\frac{dV_{31}}{dt}
    =\frac{d}{dt}\ln\left(\frac{V_{11}V_{13}^{*}}{V_{31}V_{33}^{*}} \right),\\ 
    \label{14db}
    &&\frac{1}{ V_{11}}\frac{dV_{11}}{dt} - \left(\frac{1}{
      V_{33}}\frac{dV_{33}}{dt}\right)^{*} 
  -\frac{1}{ V_{13}}\frac{dV_{13}}{dt}
    +\left(\frac{1}{ V_{31}}\frac{dV_{31}}{dt}\right)^{*}
    =\frac{d}{dt}\ln\left(\frac{V_{11}V_{31}^{*}}{V_{13}V_{33}^{*}} \right),\\ 
    \label{14dc}
    &&\frac{1}{ V_{11}}\frac{dV_{11}}{dt} + \frac{1}{
      V_{33}}\frac{dV_{33}}{dt} + \left(\frac{1}{
        V_{13}}\frac{dV_{13}}{dt}\right)^{*}
    +\left(\frac{1}{ V_{31}}\frac{dV_{31}}{dt}\right)^{*}
     =\frac{d}{dt}\ln\left(V_{11}V_{33}V_{13}^{*}V_{31}^{*}\right),
  \end{eqnarray}
\end{subequations}
we get from Eqs.~\eqref{14d} that the following functions of the CKM
matrix elements are constant during the RG evolution
\begin{equation}
    \label{14ea}
    \operatorname{Im}\left(\ln\left(\frac{V_{11}V_{13}^{*}}{V_{31}V_{33}^{*}}
    \right)\right)=
\operatorname{Im}\left(\ln\left(\frac{V_{11}V_{31}^{*}}{V_{13}V_{33}^{*}}
    \right)\right)=
\operatorname{Im}\left(\ln\left({V_{11}V_{33}}{V_{13}^{*}V_{31}^{*}}
    \right)\right)=\text{const.}
\end{equation}
We thus see that though  Eqs.~\eqref{14d} cannot be solved they give
us an information that certain functions of the CKM matrix elements do
not evolve.

The first constant is equal to the angle $\alpha$ of the unitary
triangle, the second one is equal to one of the angles of the unitary
triangle that is obtained from the multiplication of the first and
third row of the CKM matrix. The third constant is the phase of the
rephasing invariant $(V_{11}V_{33}V_{13}^{*}V_{31}^{*})$, whose
imaginary part $\operatorname{Im}(V_{11}V_{33}V_{13}^{*}V_{31}^{*})$
is the Jarlskog invariant~\cite{Jarlskog}. The physical meaning of
this constant is discussed in the next section.

\subsection{RG equations for the moduli of the CKM matrix elements}

Let us now find the RG equations for the squares of the absolute
values of the CKM matrix elements $|V_{11}|^{2}$, $|V_{33}|^{2}$,
$|V_{13}|^{2}$ and $|V_{31}|^{2}$. One can easily show that the
following relation holds
\begin{equation}
  \label{14f}
  \frac{1}{|V_{ij}|^{2}}\frac{d|V_{ij}|^{2}}{dt} = 2\operatorname{Re}
      \left(\frac{1}{V_{ij}}\frac{dV_{ij}}{dt}\right).
\end{equation}
Now, using the fact that $R^{u,d}_{ii}$ are purely imaginary we obtain
from Eqs.~\eqref{14c}
\begin{subequations}
  \label{14g}
\begin{eqnarray}
  \label{14ga}
  &&\frac{1}{ |V_{11}|^{2}}\frac{d|V_{11}|^{2}}{dt} = 
  -\frac{2}{(4\pi )^{2}}
  \alpha_{4}^{u}\left((Y_{d}^{2})_{11}-(VY_{d}^{2}V^{\dagger})_{11}\right)
 -\frac{2}{(4\pi )^{2}}
  \alpha_{2}^{d}\left((Y_{u}^{2})_{11}-(V^{\dagger}Y_{u}^{2}V)_{11}\right),\\
\label{14gb}
  &&\frac{1}{ |V_{33}|^{2}}\frac{d|V_{33}|^{2}}{dt} = 
   \frac{2}{(4\pi )^{2}}
  \alpha_{4}^{u}\left((Y_{d}^{2})_{33}-(VY_{d}^{2}V^{\dagger})_{33}\right)
 +\frac{2}{(4\pi )^{2}}
  \alpha_{2}^{d}\left((Y_{u}^{2})_{33}-(V^{\dagger}Y_{u}^{2}V)_{33}\right),\\
\label{14gc}
  &&\frac{1}{ |V_{13}|^{2}}\frac{d|V_{13}|^{2}}{dt} = 
  -\frac{2}{(4\pi )^{2}}
  \alpha_{4}^{u}\left((Y_{d}^{2})_{33}-(VY_{d}^{2}V^{\dagger})_{11}\right)
 +\frac{2}{(4\pi )^{2}}
  \alpha_{2}^{d}\left((Y_{u}^{2})_{11}-(V^{\dagger}Y_{u}^{2}V)_{33}\right),\\
\label{14gd}
  &&\frac{1}{ |V_{31}|^{2}}\frac{d|V_{31}|^{2}}{dt} = 
  \frac{2}{(4\pi )^{2}}
  \alpha_{4}^{u}\left((Y_{d}^{2})_{11}-(VY_{d}^{2}V^{\dagger})_{33}\right)
 -\frac{2}{(4\pi )^{2}}
  \alpha_{2}^{d}\left((Y_{u}^{2})_{33}-(V^{\dagger}Y_{u}^{2}V)_{11}\right).
\end{eqnarray}
\end{subequations}
These equations form the complete set of the RG equations for the CKM
matrix that do not contain unnknown functions. Eqs.~\eqref{14g} can be
rewritten in terms of the squares of the absolute values of the CKM
matrix
\begin{subequations}
\label{16}
\begin{multline}
\frac{1}{\left| {V}_{11}\right| ^{2}}\frac{d\left| {V}_{11}\right| ^{2}}{dt}
=\frac{2}{(4\pi )^{2}}\alpha _{4}^{u}\left( \left| {V}_{13}\right|
^{2}\left( Y_{b}\right) ^{2}+\left| {V}_{12}\right| ^{2}\left(
Y_{s}\right) ^{2}-\left( 1-\left| {V}_{11}\right| ^{2}\right)
\left( Y_{d}\right) ^{2}\right) \\
+\frac{2}{(4\pi )^{2}}\alpha _{2}^{d}\left( \left| {V}_{31}\right|
^{2}\left( Y_{t}\right) ^{2}+\left| {V}_{21}\right| ^{2}\left(
Y_{c}\right) ^{2}-\left( 1-\left| {V}_{11}\right| ^{2}\right) \left(
Y_{u}\right) ^{2}\right),  \label{16a}
\end{multline}
\begin{multline}
 \frac{1}{\left| {V}_{13}\right| ^{2}} \frac{d\left| {V}_{13}\right| ^{2}}{dt}
  =\frac{2}{(4\pi )^{2}}\alpha _{4}^{u}\left( -\left( 1-\left|
        {V} _{13}\right| ^{2}\right) \left( Y_{b}\right)
    ^{2}+\left| {V} _{12}\right| ^{2}\left( Y_{s}\right)
    ^{2}+\left| {V}_{11}\right| ^{2}\left( Y_{d}\right)
    ^{2}\right)  \\
  +\frac{2}{(4\pi )^{2}}\alpha _{2}^{d}\left( -\left|
      {V}_{33}\right| ^{2}\left( Y_{t}\right) ^{2}-\left|
      {V}_{23}\right| ^{2}\left( Y_{c}\right) ^{2}+\left(
      1-\left| {V}_{13}\right| ^{2}\right) \left( Y_{u}\right)
    ^{2}\right),   \label{16b}
\end{multline}
\begin{multline}
\frac{1}{ \left| {V}_{31}\right| ^{2}}\frac{d\left| {V}_{31}\right| ^{2}}{dt}
=-\frac{2}{(4\pi )^{2}}\alpha _{4}^{u}\left( \left| {V}_{33}\right|
^{2}\left( Y_{b}\right) ^{2}+\left| {V}_{32}\right| ^{2}\left(
Y_{s}\right) ^{2}-\left( 1-\left| {V}_{31}\right| ^{2}\right) \left(
Y_{d}\right) ^{2}\right)   \\
-\frac{2}{(4\pi )^{2}}\alpha _{2}^{d}\left( \left( 1-\left| {V}
_{31}\right| ^{2}\right) \left( Y_{t}\right) ^{2}-\left| {V}
_{21}\right| ^{2}\left( Y_{c}\right) ^{2}-\left| {V}_{11}\right|
^{2}\left( Y_{u}\right) ^{2},\right)
\label{16c}
\end{multline}
\begin{multline}
  \frac{1}{\left| {V}_{33}\right| ^{2}}\frac{d\left| {V}_{33}\right|
    ^{2}}{dt} =\frac{2}{(4\pi )^{2}}\alpha _{4}^{u}\left( \left(
      1-\left| {V} _{33}\right| ^{2}\right) \left( Y_{b}\right)
    ^{2}-\left| {V} _{32}\right| ^{2}\left( Y_{s}\right) ^{2}-\left|
      {V}_{31}\right|
    ^{2}\left( Y_{d}\right) ^{2}\right)\\
  +\frac{2}{(4\pi )^{2}}\alpha _{2}^{d}\left( \left( 1-\left| {V}
        _{33}\right| ^{2}\right) \left( Y_{t}\right) ^{2}-\left| {V}
      _{23}\right| ^{2}\left( Y_{c}\right) ^{2}-\left| {V}_{13}\right|
    ^{2}\left( Y_{u}\right) ^{2}\right),
  \label{16d}
\end{multline}
\end{subequations}
where we use an intuitive notation $\left( Y_{u}^{2}\right)
_{11}=Y_{u}^{2}$, $\left( Y_{u}^{2}\right) _{22}=Y_{c}^{2}$, $\left(
  Y_{u}^{2}\right) _{33}=Y_{t}^{2}$, $\left( Y_{d}^{2}\right)
_{11}=Y_{d}^{2}$, $\left( Y_{d}^{2}\right) _{22}=Y_{s}^{2}$, $\left(
  Y_{d}^{2}\right) _{33}=Y_{b}^{2}$.  We can see from these equations
that one needs to know the running of the $\left( Y_{u,d}^{2}\right)
_{ii}$ to obtain the evolution of the CKM matrix.

\section{RG equations for various models}\label{RGModels}

The hierarchy of the Yukawa couplings, Eq.~(\ref{3}) and the hierarchy
of the CKM matrix, best seen in the Wolfenstein
representation~\cite{Wolfenstein} permit to obtain the approximate,
simpler form of Eqs.~(\ref{16}), that is compatible with the one loop
approximation. 

For the approximate form of the renormalization group equations we will
consider two scenarios:
\begin{enumerate}
\item Standard Model or two Higgs extensions of the Standard Model and
  Minimal Supersymmetric Standard Model with $v_{1}\approx v_{2}$.
\item Two Higgs extension of the Standard Model and Minimal
  Supersymmetric Standard model with
  $\frac{v_{1}}{v_{2}}=\frac{m_{t}}{m_{b}}$.
\end{enumerate}
The approximation principle that we will follow consists in keeping in
the RG equations the terms of relative order order of $\lambda ^{3}$
or lower, what is also consistent with the Wolfenstein parametrization
of the CKM matrix.

In the first scenario the squares of the Yukawa couplings of the down
quarks can be neglected in comparison to those of the up quarks. Also
the squares of the Yukawa couplings of the charm and up quarks can be
neglected in comparison to the top quark Yukawa coupling and thus the
resulting renormalization group equations are considerably simplified
and read
\begin{subequations}
\label{18}
\begin{eqnarray}
  \displaystyle\frac{1}{\left| {V}_{11}\right| ^{2}}\frac{d\left|
      {V}_{11}\right| ^{2}}{dt} &\displaystyle=&\displaystyle\frac{2}{(4\pi
    )^{2}}\alpha _{2}^{d}\left| {V}_{31}\right| ^{2}Y_{t}^{2},
  \label{18a}\\
 \displaystyle \frac{1}{\left| {V}_{13}\right| ^{2}} \frac{d\left|
      {V}_{13}\right| ^{2}}{dt} &\displaystyle=&\displaystyle-\frac{2}{(4\pi
    )^{2}}\alpha _{2}^{d}\left| {V}_{33}\right| ^{2}Y_{t}^{2},
  \label{18b}\\
  \displaystyle\frac{1}{\left|{V}_{31}\right| ^{2}}\frac{d\left|
      {V}_{31}\right| ^{2}}{dt} &\displaystyle=&\displaystyle-\frac{2}{(4\pi)^{2}}\alpha
  _{2}^{d}\left( 1-\left| {V}_{31}\right|^{2}\right) Y_{t}^{2},
  \label{18c}\\ 
 \displaystyle \frac{1}{\left|
    {V}_{33}\right|^{2}}\frac{d\left| {V}_{33}\right|
  ^{2}}{dt} &\displaystyle=&\displaystyle\frac{2}{(4\pi)^{2}}\alpha _{2}^{d}\left( 1-\left|
    {V}_{33}\right| ^{2}\right) Y_{t}^{2}.  \label{18d}
\end{eqnarray}
\end{subequations}
These equations were derived earlier and were explicitly solved in
Ref.~\cite{prd_66_116007,plb_479_181} so we will not discuss them in
more detail.

The second scenario is physically very interesting if we assume that
the origin of the hierarchy of masses between the top and bottom
quarks lies in the hierarchy of the vacuum expectation values of the
Higgs fields:
\begin{equation}
  \frac{v_{1}}{v_{2}}\approx \frac{m_{t}}{m_{b}}  \label{19}
\end{equation}
from which we obtain
\begin{equation}
  \left( Y_{u}\right) _{33}\approx \left( Y_{d}\right) _{33},  \label{20}
\end{equation}
so the Yukawa couplings of the down quarks have to be kept in the
equations.  
From the hierarchy in Eq.~(\ref{21}) one derives after straightforward
calculations the following equations for the CKM matrix elements
\begin{subequations}
\label{22}
\begin{eqnarray}
  &&\frac{1}{\left| {V}_{11}\right| ^{2}}\frac{d\left| {V}_{11}\right|
    ^{2}}{dt}=\frac{2}{(4\pi )^{2}}\left( \alpha _{2}^{d}\left| {V}
      _{31}\right| ^{2}Y_{t}^{2}+\alpha _{4}^{u}\left| {V}_{13}\right|
    ^{2}Y_{b}^{2}+\alpha _{4}^{u}\left| {V}_{12}\right|
    ^{2}Y_{s}{}^{2}\right),
  \label{22a}\\
  &&\frac{1}{\left| {V}_{13}\right| ^{2}}\frac{d\left| {V}_{13}\right|
    ^{2}}{dt}=-\frac{2}{(4\pi )^{2}}\left( \alpha _{2}^{d}\left| {V}
      _{33}\right| ^{2}Y_{t}^{2}+\alpha _{4}^{u}\left( 1-\left| {V}
        _{13}\right| ^{2}\right) Y_{b}^{2}\right),   \label{22b}\\
  &&\frac{1}{\left| {V}_{31}\right| ^{2}}\frac{d\left| {V}_{31}\right|
    ^{2}}{dt}=-\frac{2}{(4\pi )^{2}}\left( \alpha _{2}^{d}\left(
      1-\left| {V}
        _{31}\right| ^{2}\right) Y_{t}{}^{2}+\alpha _{4}^{u}\left| {V}
      _{33}\right| ^{2}Y_{b}^{2}\right),
  \label{22c}\\
  &&\frac{1}{\left| {V}_{33}\right| ^{2}}\frac{d\left| {V}_{33}\right|
    ^{2}}{dt}=\frac{2}{(4\pi )^{2}}\left( \alpha _{2}^{d}Y_{t}{}^{2}+\alpha
    _{4}^{u}Y_{b}^{2}\right) \left( 1-\left| {V}_{33}\right| ^{2}\right).
\label{22d}
\end{eqnarray}
\end{subequations}
Eqs.~(\ref{18}) form a complete set of equations for the CKM evolution
for the first scenario and Eqs.~(\ref{22}) describe the CKM matrix
evolution in the second scenario. Eqs.~(\ref{18}) require the
knowledge of the evolution of the top quark Yukawa coupling $Y_{t}(t)$
while for the second scenario we need the knowledge of the Yukawa
coupling evolution of the top, bottom and strange quarks, $Y_{t}(t)$,
$Y_{b}(t)$, $Y_{s}(t)$.

\subsection{Solution for the first scenario}
Eqs.~(\ref{18}) have been analized before \cite{prd_66_116007} but we
shall make some additional comments here.  Let us notice that after
subtracting Eq.~(\ref{18c}) from Eq.~(\ref{18a}) and Eq.~(\ref{18b})
from Eq.~(\ref{18d}) we obtain
\begin{subequations}
\label{23}
\begin{eqnarray}
  \frac{1}{\left| {V}_{11}\right| ^{2}}\frac{d\left| {V}_{11}\right|
    ^{2}}{dt}-\frac{1}{\left| {V}_{31}\right| ^{2}}\frac{d\left| {V}
      _{31}\right| ^{2}}{dt}=\frac{2}{(4\pi )^{2}}\alpha _{2}^{d}Y_{t}{}^{2}
\label{23a}\\
\frac{1}{\left| {V}_{33}\right| ^{2}}\frac{d\left| {V}_{33}\right|
  ^{2}}{dt}-\frac{1}{\left| {V}_{13}\right| ^{2}}\frac{d\left| {V}
    _{13}\right| ^{2}}{dt}=\frac{2}{(4\pi )^{2}}\alpha _{2}^{d}Y_{t}{}^{2}
\label{23b}
\end{eqnarray}
\end{subequations}
Eqs.~(\ref{23}) can be easily integrated and from them it follows that
the evolution of the ratios $\frac{\left| {V}_{11}\right|
  ^{2}}{\left| {V }_{31}\right| ^{2}}$ and $\frac{\left|
    {V}_{33}\right| ^{2}}{\left| { V}_{13}\right| ^{2}}$ is
the same. Moreover, if we substract Eq.~(\ref{23b}) from
Eq.~(\ref{23a}) we get
\begin{equation}
  \frac{1}{\left| {V}_{11}\right| ^{2}}\frac{d\left| {V}_{11}\right|
    ^{2}}{dt}-\frac{1}{\left| {V}_{31}\right| ^{2}}\frac{d\left| {V}
      _{31}\right| ^{2}}{dt}-\frac{1}{\left| {V}_{33}\right| ^{2}}\frac{
    d\left| {V}_{33}\right| ^{2}}{dt}+\frac{1}{\left| {V}_{13}\right|
    ^{2}}\frac{d\left| {V}_{13}\right| ^{2}}{dt}=0  \label{24}
\end{equation}
so in the case of the first scenario the ratio
\begin{equation}
  \frac{\left| {V}_{11}\right| ^{2}}{\left| {V}_{33}\right| ^{2}}\frac{
    \left| {V}_{13}\right| ^{2}}{\left| {V}_{31}\right|
    ^{2}}=\text{const}  \label{25}
\end{equation}
is also constant.

Next from Eqs.~(\ref{18a}), (\ref{18c}) and Eqs.~(\ref{18b}),
(\ref{18d}) one can derive that the following ratios are constants
during the RG evolution
\begin{equation}
  \frac{\left| {V}_{13}\right| ^{2}}{1-\left| {V}_{33}\right| ^{2}}
  =\text{const},\quad\quad\frac{\left|
      {V}_{11}\right| ^{2}}{1-\left| {V}_{31}\right| ^{2}}
  =\text{const}.  \label{26}
\end{equation}
Eqs.~\eqref{23} and~\eqref{26} allow to derive the exact evolution of
the CKM matrix for the first scenario. From these equations it also
follows that the evolution of the CKM matrix depends only on one
function of energy.

\subsection{Discussion of the second scenario}
The analysis of the RG equations for the second scenario
Eqs.~(\ref{22}) is more involved. One can see that Eq.~(\ref{22d}),
can be rewritten in the following way
\begin{equation}
  \frac{d}{dt}\left( \ln \frac{\left| {V}_{33}\right|
      ^{2}}{1-\left| {V}_{33}\right| ^{2}}\right)
  =\frac{2}{(4\pi)^{2}}\left( \alpha_{2}^{d}Y_{t}{}^{2}+\alpha
    _{4}^{u}Y_{b}^{2}\right)   \label{27} 
\end{equation}
so the evolution of $\left| {V}_{33}\right| ^{2}$ is explicitly known.

Next, using the solution of Eq.~(\ref{27}) one can solve
Eqs.~(\ref{22b}) and~(\ref{22c}) for $\left| {V}_{13}\right| ^{2}$
and $\left| {V} _{31}\right| ^{2}$. Eq.~(\ref{22a}) can be solved
using the solutions of Eqs.~(\ref{22b}), (\ref{22c}) and (\ref{22d}).
The full phenomenological analysis of Eqs.~(\ref{22}) is beyond the
scope of this paper and will be published elsewhere.

\section{General discussion and conclusions}


We have obtained the full set of the one loop RG equations for the CKM
matrix for three models: Standard Model, two Higgs extension of the
Standard Model and the Minimal Supersymmetric Standard Model. They are
given in Eqs.~\eqref{14g} or~\eqref{16} and they form a set of
coupled, non linear ordinary differential equations for the squares of
the moduli of the CKM matrix elements. Eqs.~\eqref{14g} were derived
from the RG equations for the quark Yukawa couplings by the
elimination of the right diagonalizing matrices. One can see that
Eqs.~\eqref{14g} depend only on $\alpha_{4}^{u}$ and $\alpha_{2}^{d}$,
\textit{i.e.}, only on a subset of $\alpha_{i}^{u,d}$,
$i=1,\ldots,5$ of the model dependent parameters of the RG equations
for the Yukawa couplings, Eqs.~\eqref{5}. This means that the
evolution of the CKM matrix can test only partially the evolution of
the Yukawa couplings.

The most important result of the paper is the demonstration from
Eq.~\eqref{14ea} that the angle $\alpha$ of the unitarity triangle is
constant during the evolution. The angle $\alpha$ is also equal to the
one of the angles of the unitarity triangle obtained by the
multiplication of the first and third rows of the CKM matrix and is
also equal to the phase of the rephasing invariant
$V_{11}V_{33}V_{13}^{*}V_{31}^{*}$, whose imaginary part is equal 
to the Jarlskog invariant of the CP violation. For the general case of
the three models this is the only constant that can be obtained. On
the other hand for the first scenario, \textit{i.e.}, for the Standard
Model and its extensions with $v_{1}\approx v_{2}$, we additionaly
have the constant in Eq.~\eqref{25}, which is equal to the ratio of
two sides of the unitarity triangle that are adjacent to the angle
$\alpha$. From this it follows that during the RG evolution the
unitarity triangle is only rescaled and its shape does not change.
This means that for the first scenario it is not possible to construct
an asymptotic model with some simple, special form of the CKM matrix.

Let us observe that the constants in Eqs.~\eqref{14ea} and~\eqref{25}
can be written in terms of the modified Buras-Wolfenstein
parameters~\cite{PhysRevD.50.3433} $\bar{\rho}=(1-\lambda^{2}/2)\rho$
and $\bar{\eta}=(1-\lambda^{2}/2)\eta$ in the following way
\begin{equation}
  \label{28}
  \operatorname{Im}\left(\ln\left(\frac{V_{11}V_{13}^{*}}{V_{31}V_{33}^{*}}
    \right)\right)
  =\frac{\eta}{\rho
    -\left(1-\frac{1}{2}\lambda^{2}\right)({\rho}^{2}+{\eta}^{2})}
  =\frac{\bar{\eta}}{\bar{\rho}(1 -\bar{\rho})-\bar{\eta}^{2}}
  =\text{const.}
\end{equation}
and
\begin{equation}
  \label{29}
\left| \frac{{V}_{11} {V}_{13}^{*}}
    {{V}_{31}{V}_{33}^{*}}\right|
  =\sqrt{\frac{\bar{\rho}^{2}+\bar{\eta}^{2}}
    {(1-\bar{\rho})^{2}+\bar{\eta}^{2}}} =\text{const.} 
\end{equation}
Eq.~\eqref{28} is valid for both scenarios, so we see that a
relatively complicated function of the Wolfenstin parameters is a
constant of the RG evolution for the general case. Eq.~\eqref{29} is
valid for the first scenario so from Eqs.~\eqref{28} and~\eqref{29} it
follows that for the first scenario the parameters $\bar{\rho}$ and
$\bar{\eta}$ are constants of the RG evolution.

\begin{acknowledgments}
  We acknowledge the financial support from SNI and CONACYT.  S.R.J.W.
  and J.H.M.O.Y thank COFAA, PIFI, EDD and SIP-Proyecto 20080642,
  Instituto Polit\'{e}cnico Nacional, Mexico.
\end{acknowledgments}

\newpage
\appendix*
\section{Parameters for the various models}

\begin{table}[h]
\caption{Coefficients $\alpha_1^l$, $l=u,d$ for various models.}
\label{table2}
\bigskip
\begin{tabular}{|l|l|l|}
\hline & \text{\; SM  and  DHM} & \; MSSM \\ \hline
\;$\displaystyle\vphantom{\frac{A^{A}}{A_{A}}}\alpha _{1}^{u}(t)=$\; 
&\;$\displaystyle -(\frac{17}{20}g_{1}^{2}+\frac{9}{4}
g_{2}^{2}+8g_{3}^{2})$\; 
&\;$\displaystyle -(\frac{13}{15}g_{1}^{2}+3g_{2}^{2}+\frac{16}{3}
g_{3}^{2})$\; \\ \hline
\;$\displaystyle\vphantom{\frac{A^{A}}{A_{A}}}\alpha _{1}^{d}(t)=$ \;
&\;$\displaystyle -(\frac{1}{4}g_{1}^{2}+\frac{9}{4}
g_{2}^{2}+8g_{3}^{2})$\; 
&\;$\displaystyle -(\frac{7}{15}g_{1}^{2}+3g_{2}^{2}+\frac{16}{3}
g_{3}^{2})$\; \\ \hline
\end{tabular}
\end{table}
\begin{table}[h]
  \caption{Coefficients $\alpha_k^l$, $k=2,\ldots,5$, $l=u,d$ for
    various models.}
\label{table3}
\bigskip
\begin{tabular}{|c|c|c|c|c|}
\hline $l$ & $\displaystyle\vphantom{\frac{A^{A}}{A_{A}}} \alpha _{2}^{l}$ &
$\alpha _{3}^{l}$ & $\alpha
_{4}^{l}$ & $\alpha _{5}^{l}$ \\ \hline
\;$\displaystyle\vphantom{\frac{A^{A}}{A_{A}}}u$\; &\;
$\frac{3}{2}b$\; &\; $3$\; &\; $\frac{3}{2}c$\; &\; $3a$\;
\\ \hline
$\displaystyle\vphantom{\frac{A^{A}}{A_{A}}}d$ & $\frac{3}{2}c$ & $3a$
& $\frac{3}{2}b$ & $3$ 
\\ \hline
\end{tabular}
\end{table}
\begin{eqnarray*}
  && (a,b,c)_{\,\text{SM}}=(1,1,-1)\\
  && (a,b,c)_{\,\text{DHM}}=(0,1,\frac{1}{3})\\
  && (a,b,c)_{\,\text{MSSM}}=(0,2,\frac{2}{3})
\end{eqnarray*}


\end{document}